\documentclass[twocolumn,showpacs,aps,prl]{revtex4}
\usepackage{graphicx,amsmath,amssymb}
\usepackage{dcolumn,fancyhdr}
\usepackage{bm,natbib,mathrsfs}
\usepackage{times}
\usepackage{txfonts}
\usepackage{epstopdf}
\usepackage{latexsym}
\usepackage{epsfig}

\newcommand{\ket}[1]{\left\vert#1\right\rangle}

\newcommand{\bra}[1]{\left\langle#1\right\vert}

\newcommand{\nbar}{\overline{n}}

\newcommand{\la}{\lambda}

	\newcommand{\tr}[1]{\textrm{Tr} \left[ {#1} \right]} 

	\newcommand{\e}[1]{e^{ {#1}}} 

\begin{document}

\title{Measuring the characteristic function of  work distribution}

\author{L. Mazzola, G. De Chiara, and M. Paternostro}
\affiliation{
Centre for Theoretical Atomic, Molecular and Optical Physics, School of Mathematics and Physics, Queen's University, Belfast BT7 1NN, United Kingdom}

\date{\today}

\begin{abstract}
We propose an interferometric setting for the ancilla-assisted measurement of the characteristic function of the work distribution following a time-dependent process experienced by a quantum system. We identify how the configuration of the effective interferometer is linked to the symmetries enjoyed by the Hamiltonian ruling the process and provide the explicit form of the operations to implement in order to accomplish our task. We finally discuss two physical settings, based on hybrid opto-/electro-mechanical devices, where the theoretical proposals discussed in our work could find an experimental demonstration.    
\end{abstract}

\maketitle

{Thermodynamics is one of the pillars of natural sciences. Its principles can predict the occurrence and efficiency of complex chemical reactions and biological processes. In physics, the conduction of heat across a medium or the concept of Òarrow of timeÓ are formulated thermodynamically. In information theory, the definitions of ÒinformationÓ and ÒentropyÓ are also given in thermodynamical terms. Moreover, the tightness of the link between information and thermodynamics can be deduced from the interpretation of the landmark embodied by Landauer's  principle~\cite{landauer}.

The dexterity characterizing the current experimental control at the microscopic scale opens up tantalising questions, the most pressing being probably the following: {\it what happens to thermodynamics when we deal with the non-quasistatic dynamics of quantum systems brought  out of equilibrium?} 
An invaluable tool for the formulation of an answer in this sense has been provided with the formulation of non-equilibrium fluctuation relations and their quantum extension~\cite{campisi,Tasaki}, which has recently enabled investigations at the crossroad of quantum physics, thermodynamics, and information theory~\cite{varie}. This includes proposals for experimental quantum thermal machines~\cite{Abah}, the study of the link between fluctuation relations and critical phenomena in many-body systems~\cite{dorner,joshi},  the verification of the Jarzynski equality~\cite{Jarzynski,Huber,Ngo}, and the extension to open dynamics~\cite{CPmaps}.

The verification and use of the Jarzynski inequality~\cite{Jarzynski,CPmaps} requires the determination of the work distribution following a process undergone by a system, 
a goal that needs feasible experimental strategies. In Ref.~\cite{Huber,Heyl}, two seminal proposals have been made: Huber {\it et al.} suggested a scheme based on the performance of projective energy measurements on the trapped-ion system undergoing a process. Their method uses an ingenious ``filtering scheme"  whose implementation, unfortunately, can be of significant practical difficulty. Heyl and Kehrein~\cite{Heyl}, on the other hand, showed that optical spectra can be used to measure the work distribution of specific non-equilibrium processes. However, their method only applies to sudden quenches and is ineffective for general processes. 

In this paper we propose a way to infer the quantum statistics of a work distribution by relying on an interferometric approach that delegates the retrieval of the information we are after to routine measurements performed on a finite-size ancilla. We demonstrate that a qubit-assisted Ramsey-like scheme is effective in fully determining the characteristic function of the work distribution following a general quantum process. The latter contains the same information as the work distribution itself and can be equally used in the framework of fluctuation relations for an out-of-equilibrium configuration. We identify the relation between symmetries in the quantum process and the corresponding Ramsey interferometer. Differently from Ref.~\cite{Huber}, our scheme does not rely on a specific setting and, by delegating the retrieval of information to single-qubit measurements, bypasses the problem of energy-eigenstate projections. In quite a stark contrast with Ref.~\cite{Heyl}, our proposal is valid for any process and can be used for a vast range of physical situations (cf. Ref.~\cite{oxf} for a related analysis on a trapped ion). As an illustration, we apply it to a (micro-/nano-)mechanical oscillator coupled to a two-level system and undergoing a displacement in phase space, which is a situation of strong experimental interest. Designing viable ways to access quantum statistics of non-equilibrium processes is a significant step towards the grounding of this fascinating area and the spurring of potential ramifications in fields such as quantum control and foundations of quantum mechanics~\cite{varie,switch,noisoon}. }

\section{Quantum fluctuation relations: a brief review} 
Here we give a brief summary of the formalism that will be used throughout this work. We consider a process undergone by system $S$ and described by a Hamiltonian $\hat{\cal H}(\lambda_t)$ depending on a {\it work parameter} $\lambda_t$, which is assumed to be externally controlled.
At $t=0^-$, $S$ is in contact with a reservoir and initialised in a thermal state $\rho^{th}_{S}(\lambda_0)={\e{-\beta\hat{\cal H}(\lambda_0)}}/{{\cal Z}(\lambda_0)}$ at inverse temperature $\beta$ and work parameter $\lambda_0$ [$\mathcal{Z}(\lambda)=\textrm{Tr}{\e{-\beta\hat{\cal H}(\lambda)}}$ is the partition function]. At $t=0^+$,  we detach $S$ from the reservoir and perform a {\it process} consisting of the change of $\lambda_t$ to its final value $\lambda_\tau$.
It is convenient to decompose the Hamiltonians connected by the process as $\hat{\cal H}(\lambda_0)=\sum_n E_n(\lambda_0) \ket{n}\bra{n}$ and $\hat{\cal H}(\lambda_\tau)=\sum_m E'_m(\lambda_\tau) \ket{m}\bra{m}$,
where $(E_n,\ket{n})$ [$(E'_m,\ket{m})$] is the $n^\textrm{th}$ [$m^\textrm{th}$] eigenvalue-eigenstate pair of the initial [final] Hamiltonian. The corresponding work distribution can be written as~\cite{Tasaki}
$P_{\rightarrow}(W):=\sum_{n,m} p^0_n\;  p^\tau_{m \vert n} \delta\left[W-(E_m'-E_n)\right]$.
Here, we have introduced the probability $p^0_n$ that the system is found in state $\ket{n}$ at time $t=0$ and the conditional probability $p^\tau_{m|n}$ to find it in $\ket{m}$ at time $\tau$ if it was initially in $\ket n$ and evolved under the action of the propagator $\hat{U}_\tau$.  $P_\to(W)$ encompasses the statistics of the initial state (given by $p_n^0$) and the fluctuations arising from quantum measurement statistics (given by $p^\tau_{m \vert n}$). For our purposes, it is convenient to define the characteristic function of $P_\to(W)$~\cite{lutz}
\begin{equation}
\chi(u,\tau)=\int\!{dW}\e{iuW}P_{\rightarrow}(W)=\tr{U^\dag_\tau\e{iu\hat{\cal H}(\lambda_\tau)}\hat U_\tau\e{-iu\hat{\cal H}(\lambda_0)}\rho^{th}_\textrm{S}(\lambda_0)}.
\label{characteristicfunction}
\end{equation}
From Eq.~\eqref{characteristicfunction}, the Jarzynski equality~\cite{Jarzynski} is found as 
$\chi(i\beta,\tau)= \langle \e{-\beta W} \rangle=
\e{-\beta \Delta F}$. The characteristic function is also crucial for the Tasaki-Crooks relation $\Delta F=(1/\beta)\ln[\chi'(v,\tau)/\chi(u,\tau)]$~\cite{Crooks,Tasaki} with $\chi'(v,\tau)$ the characteristic function of the backward process obtained taking $\lambda_\tau\to\lambda_0$ and evolving $\rho^{th}_S(\lambda_{\tau})$ through $U^\dag_\tau$). Here $\Delta{F}$ is the net change in the equilibrium free-energy of $S$. This demonstrates the central role played by the characteristic function in determining the equilibrium properties of a system. We shall now illustrate a protocol for the interferometric determination of $\chi(u,\tau)$. This would then enable the convenient evaluation of the figures of merit discussed above.

\section{A simple illustrative case} 
To fix the ideas before attacking the general protocol 
we consider the Hamiltonian for $S$ 
$\hat{\cal H}_S(t)=g(\la_t) \hat{h}$,
with $\hat h$ an operatorial part that remains unchanged irrespective of the process responsible for the change of the work parameter and specified by the function $g(\la_t)$. Clearly $\hat{\cal H}_S(t)$ commutes with itself and $\hat U_\tau=e^{-i \hat{h} \int_0^\tau g(\la_t) dt}$ at all instants of time. That is 
$[\hat{\cal H}_i,\hat{\cal H}_f]=\quad[\hat U_\tau,\hat{\cal H}_{i(f)}]=0$
with $\hat{\cal H}_i\equiv\hat{\cal H}_S(0)=g(\la_0)\hat{h}$ and $\hat{\cal H}_f\equiv\hat{\cal H}_S(\tau)=g(\la_{\tau})\hat h$. The characteristic function thus simplifies as
\begin{equation}
\begin{aligned}
\chi_{\rm s}(u)=
\mathrm{Tr}\left[e^{i (\hat{\cal H}_f-\hat{\cal H}_i)u}\rho^{th}_S(\la_0)\right]
\end{aligned}
\end{equation}
and is fully determined by the changes induced in $\hat{\cal H}_S(t)$  by the process. This allows us to make a significant progress in the illustration of our scheme. Indeed, let us introduce an ancilla qubit $A$, whose role is to assist in the measurement of $\chi_{\rm s}(u)$. Moreover, we consider the $S$-$A$ evolution $\hat G(u)\hat V(u)$, where $\hat V(u)=\e{-i\hat {\cal H}_iu}\otimes\hat\openone_A$ is a local transformation on $S$ and $\hat G(u)$ is the controlled $A$-$S$ gate 
\begin{equation}
\label{simple}
\hat G(u)=\hat \openone_S\otimes\ket{0}\!\bra{0}_A+e^{-i (\hat{\cal H}_f-\hat{\cal H}_i) u}\otimes\ket{1}\!\bra{1}_A,
\end{equation}
which applies $\e{-i (\hat{\cal H}_f-\hat{\cal H}_i) u}$ to the state of $S$ only when $A$ is in $\ket{1}_A$ and leaves it unaffected otherwise. Gates having the form $\openone_S\otimes\ket{0}\!\bra{0}_A+\hat{\mathscr{U}}_S\otimes\ket{1}\!\bra{1}_A$ (with $\hat{\mathscr{U}}_S$ a unitary for the system), which are clearly of the form of Eq.~\eqref{simple} can be generated, for instance, by $S$-$A$ Hamiltonians having the structure ${\mathscr{O}}_S\otimes\ket{1}\bra{1}_A$, with ${\mathscr{O}}_S$ an appropriate Hamiltonian term.  

\begin{figure}[!t]
\hskip-1cm{\bf (a)}\hskip4.2cm{\bf (b)}
\centering
\includegraphics[width=0.4\linewidth]{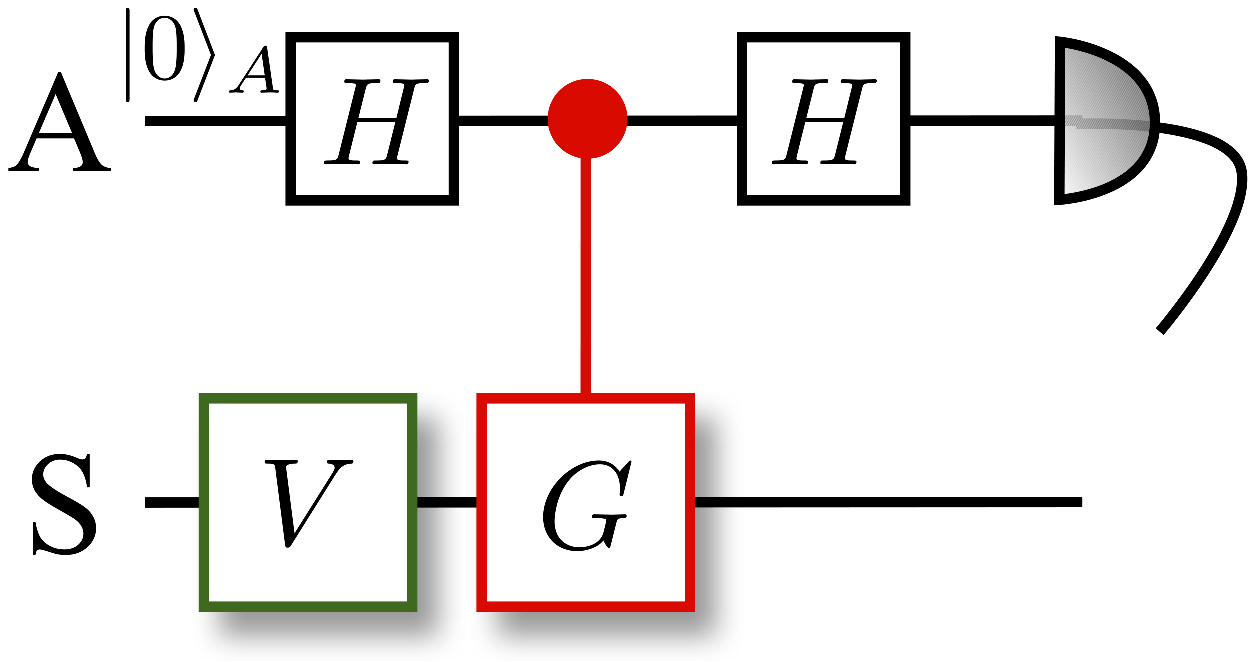}~~~~~~~\includegraphics[width=0.55\linewidth]{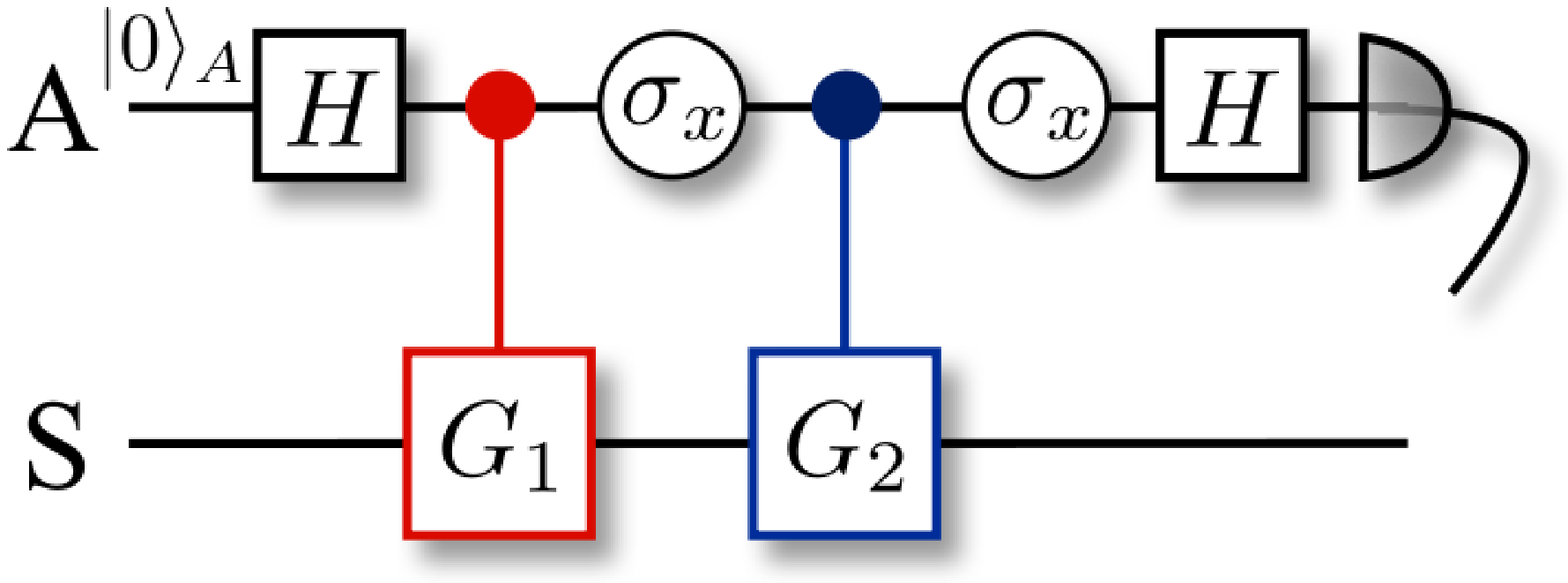}
\caption{(Color online) {\bf (a)} Quantum circuit illustrating the protocol for the measurement of $\chi_{\rm s}(u)$. The ancilla $A$ is a qubit initialised in $\ket{0}_A$ and undergoing a Hadamard gate $\hat H$. System $S$ is prepared in a thermal state $\rho^{th}_S$ and is subjected to the local transformation $\hat V$. See the body of the manuscript for the explicit form of the gates (whose dependence on  $u$ has been omitted here). {\bf (b)} Quantum circuit illustrating the scheme 
for the most general process undergone by $S$. In both panels we show the symbol for conditional $A$-$S$ gates controlled by the state of the ancilla. In panel {\bf (b)} we also picture the symbol for a full inversion gate as given by $\hat\sigma_x,A$} 
\label{schemasemplice}
\end{figure}

Inspired by Ramsey-like schemes for parameter estimation~\cite{ramsey,catastrophe}, our protocol proceeds as follows: We prepare $\ket{0}_A$ and apply a Hadamard transform $\hat H_A=(\hat\sigma_{x,A}+\hat\sigma_{z,A})/\sqrt2$~\cite{Nielsen} that changes it into the eigenstate of the $x$-Pauli matrix $\ket{+}_A=(\ket{0}_A+\ket{1}_A)/\sqrt{2}$. We then apply $\hat G(u)\hat V(u)$ on $\rho^{th}_S\otimes\ket{+}\bra{+}_A$ and subject $A$ to a second Hadamard transform [cf. Fig.~\ref{schemasemplice} {\bf (a)}]. Gate $\hat G(u)$ establishes quantum correlations between $A$ and $S$ as shown by the fact that information on $S$ can be retrieved from the ancilla as 
\begin{equation}
\label{reduced}
\begin{aligned}
\rho_A&=\textrm{Tr}_S[\hat H_A\hat{G}(u)\hat V(u)(\rho^{th}_S\otimes\ket{+}\bra{+}_A)\hat V^\dag(u)\hat{G}^\dag(u)\hat H_A]\\
&=(\hat\openone_A+\alpha\hat\sigma_{z,A}+\nu\hat\sigma_{y,A})/2
\end{aligned}
\end{equation}
with $\alpha=\textrm{Re}\chi_{\rm s}$ and $\nu=\textrm{Im}\chi_{\rm s}$. This proves the effectiveness of our protocol for the measurement of $\chi_{\rm s}(u)$, which is achieved by measuring the (experimentally straightforward) longitudinal and transverse magnetization $\langle\hat\sigma_{z,A}\rangle$ and $\langle\hat\sigma_{y,A}\rangle$ of $A$. 

\section{General protocol}
We now relax the previous assumption on the form of the Hamiltonian and consider the general case where 
$[\hat{\cal H}_i,\hat{\cal H}_f]\neq 0$ and $[\hat U_\tau,\hat{\cal H}_{i(f)}]\neq0$.
Correspondingly, the characteristic function takes the form in Eq.~(\ref{characteristicfunction}) and the interferometric approach illustrated above still applies, the only difference being the form of the controlled operation to be applied on the $S$ state. Explicitly, we should implement   
\begin{equation}
\hat G(u,\tau)=\hat U_\tau e^{-i \hat{\cal H}_i u}\otimes\ket{0}\!\bra{0}_A+e^{-i \hat{\cal H}_f u}\hat U_\tau\otimes\ket{1}\!\bra{1}_A,
\end{equation}
which can be decomposed into local transformations and $A$-controlled gates as $\hat{G}(u,\tau)=(\openone_S\otimes\hat\sigma_{x,A})\hat{G}_2(u,\tau)(\openone_S\otimes\hat\sigma_{x,A})\hat G_1(u,\tau)$ [cf. Fig.~\ref{schemasemplice} {\bf (b)}] with 
\begin{equation}
\label{sequence}
\begin{aligned}
\hat G_{1}(u,\tau)&=\hat\openone_S\otimes\ket{0}\!\bra{0}_A+\e{-i\hat{\cal H}_fu}\hat{U}_\tau\otimes\ket{1}\!\bra{1}_A,\\
\hat G_{2}(u,\tau)&=\hat\openone_S\otimes\ket{0}\!\bra{0}_A+\hat{U}_\tau\e{-i\hat{\cal H}_iu}\otimes\ket{1}\!\bra{1}_A.
\end{aligned}
\end{equation}
Using the same preparation of $A$ as above and the Hadamard transforms, we obtain a reduced state identical to the second line of Eq.~\eqref{reduced} with $\alpha\to\textrm{Re}\chi(u,\tau)$ and $\nu\to\textrm{Im}\chi(u,\tau)$. 

\section{Physical examples} 
Two situations of current experimental interest can be used to illustrate our main findings. They are both based on the hybrid coupling between a two-level system and a mechanical oscillator, which can be either microscopic (in a cavity optomechanics setup) or nanoscopic (as in electromechanics). We now show how to achieve the Hamiltonians regulating the processes that we have so far described in both scenarios and illustrate the principles of our proposal by calculating the corresponding characteristic function.  

\begin{figure}[!b]
\centering{\bf (a)}\\
\includegraphics[width=0.85\linewidth]{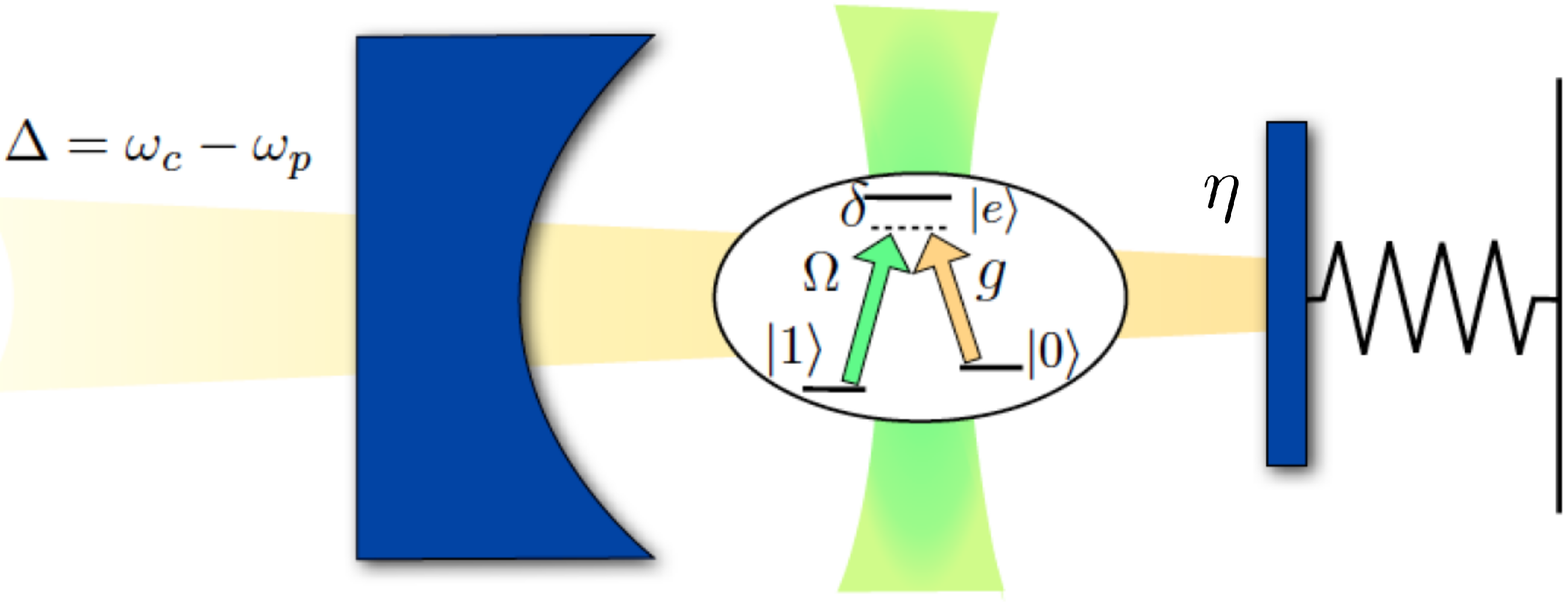}\\
\hskip-0.8cm{\bf (b)}\hskip4cm{\bf (c)}\\
\includegraphics[width=0.37\linewidth]{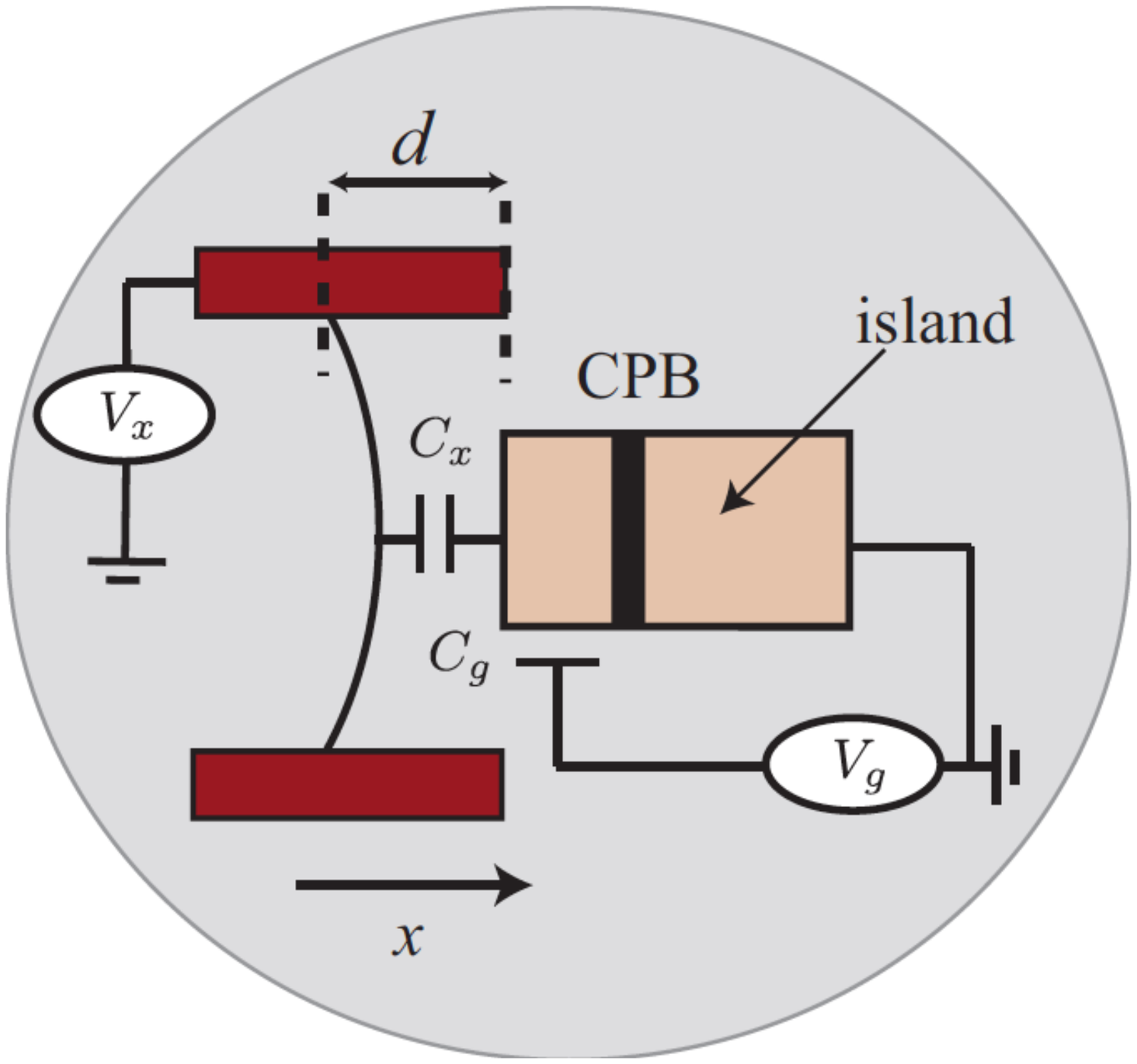}\includegraphics[width=0.6\linewidth]{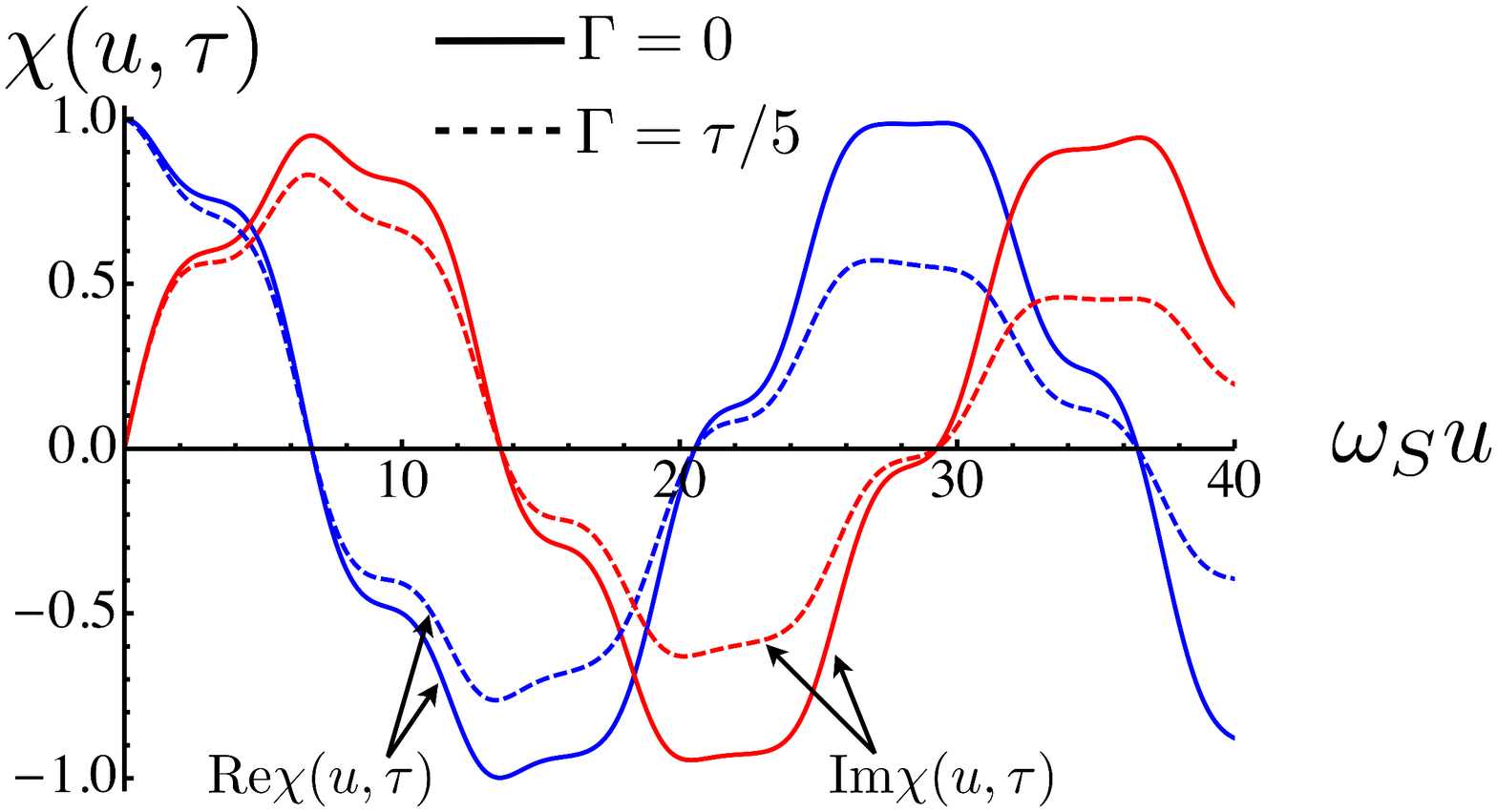}
\caption{(Color online) {\bf (a)} Hybrid micro-optomechanical setting for the measurement of $\chi(u,\tau)$. The process is undergone by a system embodied by the movable cavity mirror. The ancilla is encoded in the ground-state doublet of a three-level atom. {\bf (b)} Nano-mechanical version of the setup. System $S$ is an electrically driven nano beam (bias voltage
$V_x$). The ancilla is a CPB coupled to $S$ via the capacitance $C_x$. The state of the CPB is controlled by the gate voltage $V_g$ (coupled to the box through the capacitance $C_g$) and the Josephson energy $E_J$. {\bf (c)} Plot of $\chi(u,\tau)$ against $\omega_Su$ for $\nbar=1$, $\lambda_t=0.1\omega_S\tanh(\omega_St)$, and $\tau=10\omega^{-1}_S$. The solid [dashed] lines show real and imaginary part of the ideal [damped with $\Gamma=5\tau^{-1}$] characteristic function.} 
\label{esempi} 
\end{figure}

We start from a microscopic setting where a three-level atom in a $\Lambda$ configuration is coupled to a single-mode cavity having a movable mirror and pumped by a laser at frequency $\omega_p$. The atom is driven by a second field (frequency $\omega_i$) entering the cavity radially [cf. Fig.~\ref{esempi} {\bf (a)}]. The logical states  $\{\ket{0},\ket{1}\}$ of $A$ are encoded in the fundamental atomic doublet ($\ket{e}$ being the common excited state). The scheme includes the driving (at rate $\Omega$) of the transition $|1\rangle{\leftrightarrow}|e\rangle$ by the field at frequency $\omega_i$. The transition $|0\rangle{\leftrightarrow}|e\rangle$ is guided by the cavity field (frequency  $\omega_c$) at rate $g$. Both the fields are detuned by $\delta$ from $\ket{e}$ and we introduce the detuning 
$\Delta{=}\omega_c{-}\omega_p$. System $S$ is embodied by the movable mirror, oscillating harmonically at frequency $\omega_S$ and driven (at rate $\eta$) by the cavity through radiation-pressure ~\cite{reviewsOpt}. We assume large single-photon Raman detuning and negligible decay 
from the atomic excited state, so that 
an off-resonant two-photon Raman transition is realized (dephasing will be discussed later). We take $\Delta\gg(g,\eta)$ so that both $\ket{e}$ and the cavity field are virtually populated and can be eliminated from the dynamics. We then move to a rotating frame defined by the operator $\omega_p\hat{c}^\dag\hat{c}+\omega_i\ket{e}\!\bra{e}+\omega_{10}\ket{0}\!\bra{0}_{A}$ (we assume $\hbar{=}1$ throughout the paper) with $(\hat c,\hat c^\dag)$ the operators of the cavity field. 

We thus get $\hat{\cal H}_{\rm micro} =\omega_S\hat b^\dag\hat b+\lambda(\hat{b}^\dag+\hat{b})\otimes\ket{1}\bra{1}_{A}$ with $\lambda={\eta{g}^2\Omega^2}/{\delta^2\Delta^2}$ and $(\hat b,\hat b^\dag)$ the operators of the mechanical oscillator~\cite{vacanti}. 
Through the two-photon Raman transition, the virtual quanta resulting from the atom-cavity field interaction are transferred (by the cavity field) to $S$. The state of the latter is correspondingly displaced in phase space, in a way controlled by the state of $A$. By driving the  cavity with a bichromatic pump with frequencies $\omega_{p}\pm\omega_S/2$ and relative phase $\phi$, the effective coupling between $A$ and $S$ becomes such that displacements in any direction of the phase space of the movable mirror can be arranged~\cite{cam,noiGP,Leibfried}. This includes the possibility to fully invert the sign of $\lambda$ by arranging for $\phi=\pi$. 
Moreover, considering a time-dependent amplitude of the driving field, we get $\lambda\to\lambda_t={\eta{g}^2\Omega^2(t)}/{\delta^2\Delta^2}$, so that we finally obtain
\begin{equation}
\label{model1}
\hat{\cal H}'_{\rm micro}(t) =\omega_S\hat b^\dag\hat b+\lambda_t(\hat{b}^\dag e^{i\phi}+\hat b e^{-i\phi})\otimes\ket{1}\bra{1}_{A}.
\end{equation}
The state of $A$ can be manipulated and reconstructed through an optical probe and standard tools in quantum optics. Current progresses in the fabrication of mechanical oscillators allow for very small decoherence rates, while optical cavities with large quality factors are used in optomechanical experiments~\cite{reviewsOpt}, thus making a quasi-unitary picture plausible. However, in order to provide a full assessment of the feasibility of our scheme, we will soon provide a discussion of decoherence effects. 

A similar effective model is obtained by considering the system shown in Fig.~\ref{esempi} {\bf (b)}, which involves a nanomechanical oscillator (a {\it nano beam}) coupled capacitively to a Cooper-pair box (CPB) operating in the charge-qubit regime at the so-called charge degeneracy point~\cite{schon}. In such conditions, the dynamics of the CPB can be approximated to that of a two-level system encoded in the space spanned by states $\ket{a_\pm}$, which are symmetric and antisymmetric superpositions of states with exactly $0$ and $1$ excess Cooper
pairs in the superconducting island shown in Fig.~\ref{esempi} {\bf (b)}, and encode our ancilla. The natural Hamiltonian of the system reads 
$\hat{\cal H}_1={[\hat Q-Q_g(t)]^2}/({2C_{t}})-E_J(\ket{a_+}\bra{a_+}-\ket{a_-}\bra{a_-})+\omega_S\hat b^\dag\hat b$ with $\hat Q$ the canonical charge operator of
the CPB, $C_t$ the capacitance of the  island, $Q_g(t)=C_gV_g(t)+C_xV_x(t)$
the gate charge, $E_J$ the Josephson energy, $\omega_S$ the
frequency of the oscillator (as before)~\cite{schon}, and $V_{g[x]}$ the gate [drive] voltage. For a charge qubit at the degeneracy point, an external magnetic flux can set the conditions for negligible Josephson energy with respect to the other rates of the Hamiltonian~\cite{schon}. By defining $\hat\Sigma_{x,A}=\ket{a_+}\bra{a_-}+\ket{a_-}\bra{a_+}$, expanding
$\hat{\cal H}_1$ in series of the ratio between the actual position of
the oscillator and its equilibrium distance from the CPB (the
amplitude of the oscillations is assumed small enough that only first-order
terms are retained) and
adjusting the voltages so that $Q_g(t)\simeq0$, the Hamiltonian of the system becomes 
$\hat{\cal H}_{\rm nano}(t)=\omega_S\hat b^\dag\hat b+\lambda_t(\hat b+\hat b^\dag)\otimes\hat\Sigma_{x,A}$
 (the form of $\lambda_t$ in this case is inessential for our tasks)~\cite{rabl,nota}. The state of $A$ can be processed (measured) tuning $V_g(t)$ (using single-electron transistors)~\cite{schon}.

Both models describe a harmonic oscillator driven by an external force that depends on the state of the ancilla. From now on, in order to fix the ideas, we concentrate on the model embodied by Eq.~\eqref{model1}. The process that we aim ad discussing here is embodied by a rapid change $\lambda_0=0\to\lambda_\tau$ in the work parameter entering the system's Hamiltonian $\hat{\cal H}_{\rm osc}(t)=\omega_S\hat b^\dag\hat b+\lambda_t(\hat b+\hat b^\dag)$, which implements a displacement of the state of $S$ in its associated phase-space. The fact that, contrary to our assumptions so far, $A$ conditions only the term $\lambda_t(\hat b+\hat b^\dag)$ in $\hat{\cal H}'_{\rm micro}(t)$ and not the whole $\hat{\cal H}_{\rm osc}(t)$ results in gates $\hat{\tilde G}(u,\tau)$ and $\hat{\tilde G}_{1,2}$ that are slightly different from those given in Eq.~(\ref{sequence}). However, a detailed calculation shows that such differences are inessential to the effectiveness of the proposed protocol. While we refer to the Appendix for a rigorous and detailed analysis, for the sake of completeness here we provide a brief account of the form of such conditional gates. More specifically, the reconstruction of the $\chi(u,\tau)$ associated with the process at hand is possible using the conditional gate $\hat{\tilde G}(u,\tau)=(\openone_S\otimes\hat\sigma_{x,A})\hat{\tilde G}_2(u,\tau)(\openone_S\otimes\hat\sigma_{x,A})\hat{\tilde G}_1(u,\tau)$ with
$\hat{\tilde G}_1(u,\tau)=\hat{\cal G}(u)\hat{\cal K}(\tau)e^{i\hat{\cal H}_{\rm free}\tau}$ and $\hat{\tilde G}_2(u,\tau)=\hat{\cal K}(\tau)e^{i\hat{\cal H}_{\rm free}\tau}$. Here $\hat{\cal H}_{\rm free}=\omega_S\hat b^\dag\hat b$,  $\hat{\cal K}(\tau)=\hat{\cal T}e^{-i\int^\tau_0\hat{\cal H}'_{\rm micro}(t)dt}$ (in the Appendix we give the explicit form of such gate), $\hat{\cal T}$ is the time-ordering operator, and
\begin{equation}
\hat{\cal G}(u)\equiv e^{-i \hat{\cal H}'_{\rm micro}(\tau) u}=e^{-i \hat{\cal H}_{\rm free} u}\ket{0}\bra{0}_A+
e^{-i \hat{\cal H}_{\rm osc}(\tau) u}\ket{1}\bra{1}_A,
\end{equation}
which is obtained by setting the work parameter to its final value $\lambda_\tau$ and evolving for a time $u$.
A calculation based on phase-space methods allows us to  evaluate the state of $A$ associated with the process. Following our protocol and using values of the parameters in typical ranges for the suggested microscopic experimental scenario~\cite{vacanti}, an initial thermal state of mean occupation number $\nbar$, and a rapid change of $\lambda_\tau$, we find
the behavior of $\chi(u,\tau)$ shown in Fig.~\ref{esempi} {\bf (c)}.

 Let us now briefly assess the case embodied by $\hat{\cal H}_{\text{nano}}(t)$. This differs from the one illustrated above due to the fact that, differently from $\hat{\cal H}'_{\rm micro}(t)$, the $\hat{\Sigma}_{x,A}$ operator enters the coupling with the system. In principle, this makes the implementation of our protocol different from the micro mechanical case. However, as illustrated in   the Appendix, such differences can be removed using local operations applied to the CPB and the nano beam independently. This means that the Hamiltonian  for the nanomechanical configuration can be turned into a model formally equivalent to $\hat{\cal H}'_{\rm micro}(t)$, thus enabling the use of the same gates identified above without the needs to re-design the whole protocol [cf. the Appendix for a formal proof].

To evaluate the feasibility of our proposal, it is important to consider the effect of decoherence. The most critical influence would come from dephasing affecting the quantum coherences in the $A$ state, which are key to the success of our protocol. This can be easily included in our analysis by considering an exponential decay (at rate $\Gamma$) of the off-diagonal elements of the state of $A$ between the gates $\hat G_{1,2}$ (we assume that local rotations are performed so quickly that no detrimental effect would be observed). This results in the decay of $\chi(u,\tau)$, as shown in Fig.~\ref{esempi} {\bf (c)}, where quite a large damping rate is considered. Yet, the features of the characteristic function remain fully revealable. A different analysis holds for a decoherence-affected process undergone by the system. As already discussed, this requires a redefinition of $\chi(u,\tau)$ in terms of Kraus operators, as recently shown by Albash {\it et al.} in~\cite{Ngo}. Our preliminary assessment shows that the general working principles of our interferometric scheme hold unchanged even in this case. A full analysis will be presented in the Appendix. 

\section{Conclusions}
We have proposed an interferometric protocol for the measurement of the characteristic function of the work distribution corresponding to a process enforced on a system. The scheme requires both local and $A$-controlled operations on $S$, and shares similarities with Ramsey-based strategies for parameter estimation. Albeit our proposal bears no dependence on a specific experimental setting and is  applicable to any system allowing for a controllable system-ancilla interaction and the agile measurement of $A$~\cite{oxf}, we have illustrated it discussing the case of a mechanical oscillator undergoing a phase-space displacement and coupled to an ancilla. This embodies an interesting out-of-equilibrium quantum dynamics of current strong experimental interest. As $\chi(u,\tau)$ is a key element in the framework of quantum fluctuation relations, designing viable strategies for its inference is an important step forward for the grounding of out-of-equilibrium quantum thermodynamics. Our proposal contributes to such a quest by opening up the possibility for an experimental verification of the connections between out-of-equilibrium quantum statistics and criticality in a quantum many-body system~\cite{dorner,catastrophe,switch}. Interesting routes for the application of our protocol include the study of the properties of quantum thermal machines~\cite{noisoon}.

\noindent
\section{Acknowledgments} We are grateful to R. Dorner, J. Goold, K. Modi, F. L Semi${\rm\tilde{a}}$o, R. M. Serra, D. Soares-Pinto, and V. Vedral for invaluable discussions. LM and MP thank the Universidade Federal do ABC, Sao Paulo (Brazil) for hospitality during the completion of this work. LM is supported by the EU through a Marie Curie IEF Fellowship. MP thanks the UK EPSRC for a Career Acceleration Fellowship and a grant awarded under the ``New Directions for Research Leaders" initiative (EP/G004579/1).

\section{Appendix}

In this Appendix we address in details the examples provided in the main text, showing how to construct the gates required by our interferometric 
proposal in both the micromechanical and nanomechanical case. 

\maketitle
\renewcommand{\theequation}{A-\arabic{equation}}
\setcounter{equation}{0}  
\subsection{Micromechanical system}
Here we demonstrate how, using the Hamiltonian presented in Eq.~(7) of the main text, 
we can generate all the gates needed to reconstruct the characteristic function of work of a mechanical harmonic oscillator undergoing a process embodied by   
\begin{equation}  
\begin{aligned}
\hat{\cal H}_{\rm free}=\omega_S\hat b^\dag\hat b\to\hat{\cal{H}}_{\rm osc}(t)=\hat{\cal H}_{\rm free}+\lambda_t (\hat{b}^\dag e^{i\phi}+\hat{b} e^{-i\phi}).
\end{aligned}
\end{equation}
Physically, as the work parameter is changed from $\lambda_0=0$ to $\lambda_\tau$, the harmonic oscillator is displaced in its phase space. Without affecting the generality of our protocol, we set $\phi=0$. At the start of the process, the harmonic oscillator is prepared in the thermal state $\rho^{th}_{S}(0)$ at inverse temperature $\beta$ (cf. main text). As stated in the main text, the conditional gate $\hat G(u,\tau)$ needed to implement our scheme can be decomposed as 
\begin{equation}
\hat{G}(u,\tau)=(\openone_S\otimes\hat\sigma_{x,A})\hat{G}_2(u,\tau)(\openone_S\otimes\hat\sigma_{x,A})\hat G_1(u,\tau)
\end{equation}
with 
\begin{equation}
\label{sequence}
\begin{aligned}
\hat G_{1}(u,\tau)&=\hat\openone_S\otimes\ket{0}\!\bra{0}_A+\e{-i\hat{\cal H}_fu}\hat{U}_\tau\otimes\ket{1}\!\bra{1}_A,\\
\hat G_{2}(u,\tau)&=\hat\openone_S\otimes\ket{0}\!\bra{0}_A+\hat{U}_\tau\e{-i\hat{\cal H}_iu}\otimes\ket{1}\!\bra{1}_A.
\end{aligned}
\end{equation}
However, as pointed out in the main text, in our example the ancilla controls only the system's term $\lambda_t(\hat b^\dag+\hat b)$ rather than the full Hamiltonian  $\hat{\cal H}_{\rm osc}(t)$. This implies some slight changes to the form of the gates $\hat{G}(u,\tau)$ and $\hat G_{1,2}(u,\tau)$ given in Eq.~(6) of the main text and reported above. Consistently with the presentation given in the main text, we label such gates as $\hat{\tilde G}(u,\tau)$ and $\hat{\tilde G}_{1,2}(u,\tau)$. Here we show how, in turn, $\hat{\tilde G}_{1,2}(u,\tau)$ can be decomposed in gates that are directly generated by either the free evolution of the system or the joint evolution of system and ancilla.
We now introduce 
\begin{equation}\begin{split}
 \hat{\cal K}(\tau)&=\hat{\cal T}e^{-i \int_0^\tau \hat{\cal H}'_{\rm micro}(t) dt}\\
&=e^{-i \hat{\cal H}_{\rm free} \tau}\ket{0}\bra{0}_A+\hat{\cal T}
 e^{-i \int_0^\tau \hat{\cal H}_{\rm osc}(t) dt}\ket{1}\bra{1}_A
\end{split}\end{equation}
with $\hat{\cal H}'_{\rm micro}(t)$ defined in Eq.~(7) of the main text and $\hat{\cal T}$ the time-ordering operator. 
In Ref.~\cite{GZ} it is shown that 
$\hat U_\tau\equiv \hat{\cal T}e^{-i \int_0^\tau \hat{\cal H}_{\rm osc}(t) dt}=\hat D(\alpha_\tau) e^{-i \hat{\cal H}_{\rm free} \tau},$ 
where $\hat D(\alpha_\tau)=\exp[{\hat b^\dag \alpha_\tau-\hat b\alpha^*_\tau}+i \epsilon(\tau)]$ is a displacement operator with amplitude $\alpha_\tau=-i e^{-i \omega_S\tau}\int_0^\tau \lambda_t e^{i \omega_S t} dt$ and $e^{i \epsilon(\tau)}$ is an inessential phase factor (that cancels out during the calculations). Therefore, $\hat{\cal K}(\tau)$ can be recast into
\begin{equation}\begin{split}
\hat{\cal K}(\tau)=\left(\ket{0}\bra{0}_A+\hat D(\alpha_\tau) \otimes\ket{1}\bra{1}_A\right)e^{-i \hat{\cal H}_{\rm free} \tau}.
\end{split}\end{equation}
We introduce also the gate $\hat{\cal G}(u)$, which is obtained by setting the value of the work parameter to its final value $\lambda_\tau$ at time $\tau$ and letting $S$ and $A$ evolve for a time $u$. That is
\begin{equation}\begin{split}
&\hat{\cal G}(u)\equiv e^{-i \hat{\cal H}'_{\rm micro}(\tau) u}=e^{-i \hat{\cal H}_{\rm free} u}\ket{0}\bra{0}_A+
e^{-i \hat{\cal H}_{osc}(\tau) u}\ket{1}\bra{1}_A.
\end{split}\end{equation}
With these at hand, we build the gates needed for our task as
\begin{equation}
\hat{\tilde G}_1(u,\tau)=\hat{\cal G}(u)\hat{\cal K}(\tau)e^{i \hat{\cal{H}}_{\rm free}\tau},~~~\hat{\tilde G}_2(u,\tau)=\hat{\cal K}(\tau)e^{i \hat{\cal{H}}_{\rm free}\tau}.
\end{equation}
Here, the {\it inverse-time} free evolutions ruled by $\hat{\cal H}_{\rm free}$ are implemented using the well-known identity $e^{i \hat{\cal H}_{\rm free}\tau}=e^{-i \hat{\cal H}_{\rm free}(2\pi/\omega_S-\tau)}$~\cite{Lloyd}.
Combining such results we find  
\begin{equation}
\begin{aligned}
\hat{\tilde G}(u,\tau)&=(\openone_S\otimes\hat\sigma_{x,A})\hat{\tilde G}_2(u,\tau)(\openone_S\otimes\hat\sigma_{x,A})\hat{\tilde G}_1(u,\tau)\\
&=\hat{D}(\alpha_\tau) e^{-i \hat{\cal H}_i u}\otimes\ket{0}\!\bra{0}_A+e^{-i \hat{\cal H}_f u}\hat{D}(\alpha_\tau)\otimes\ket{1}\!\bra{1}_A,
\end{aligned}
\end{equation}
where $\hat{\cal H}_i=\hat{\cal H}_{\rm free}$ and $\hat{\cal H}_f=\hat{\cal H}_{\rm osc}(\tau)$. Although this expression is not identical to Eq.~(5) in the main text, it allows us to reconstruct the characteristic function of the stated process and is thus absolutely equivalent to it, as far as our protocol is concerned, as we demonstrate in what follows.

Let us use $\hat{\tilde G}(u,\tau)$  in our protocol for the measurement of the characteristic function $\chi(u,\tau)$. For the sake of argument we explicitly consider the reduced density matrix of the ancilla $\rho'_A=\textrm{Tr}_S[\hat{\tilde G}(u,\tau)(\rho^{th}_S\otimes\ket{+}\bra{+}_A)\hat{\tilde G}^\dag(u,\tau)]$. This differs from the form discussed in the main text only for the application of the second Hadamard gate on $A$ and is thus locally equivalent to it. After a straightforward calculation we find
\begin{equation}\begin{split}
\rho'_A&= \frac{1}{2}\openone_A+\frac{1}{2}[f(u,\tau)\ket{0}\bra{1}_A+h.c.]
\end{split}\end{equation}
with $f(u,\tau)=\textrm{Tr}_S[\hat D(\alpha_\tau)e^{-i \hat{\cal H}_i u}\rho^{th}_S(0) \hat D^\dag(\alpha_\tau)e^{i \hat{\cal H}_f u}]$.
We now show that $f(u,\tau)$ is exactly the characteristic function of the considered process. To this end, it is enough to introduce the identity operator $\openone_S=e^{-i \hat{\cal H}_{i}\tau}e^{i \hat{\cal H}_{i}\tau}$ at the right and left of $\rho^{th}_S(0)$. As the latter is thermal, it is invariant under the action of the free evolution, and we can write
\begin{equation}\begin{split}
f(u,\tau)&=\textrm{Tr}_S[\hat D(\alpha_\tau)e^{-i \hat{\cal H}_{i} \tau}e^{-i \hat{\cal H}_i u}\rho^{th}_{\rm S} e^{i \hat{\cal H}_i\tau}\hat D^\dag(\alpha_\tau)e^{i \hat{\cal H}_f u}]\\
&=\textrm{Tr}_S[\hat U_\tau e^{-i \hat{\cal H}_i u}\rho^{th}_S \hat U_\tau e^{i \hat{\cal H}_f u}]\equiv \chi(u,\tau).
\end{split}\end{equation}
We have thus recovered the full expression for $\chi(u,\tau)$. By applying now the second Hadamard gate to $\rho'_A$, we recover the second line of Eq. (4) in the main text, thus concluding our demonstration.

\subsection{Nanomechanical system}

In the limit of validity of the Hamiltonian $\hat{\cal H}_{\rm nano}$ given in the main text, the formal difference between the example drawn in the nanomechanical domain and $\hat{\cal H}'_{\rm micro}$  is in the form of the operator for subsystem $A$. However, it is straightforward to show that the two models are equivalent and the same gate decomposition given above can be used. To see this, it is enough to first add an extra nano-beam term of the form $\lambda_t(\hat b^\dag+\hat b)$ to the Hamiltonian $\hat{\cal H}_1$ defined in the main text. This can be done by adding a voltage to an extra lead placed close to the nano beam and opposite to the CPB in the setup shown in Fig. 2 {\bf (b)} of the main text. We then consider the unitarily transformed Hamiltonian $\hat{\cal H}'_{\rm nano}(t)=(\openone_S\otimes\hat{\tilde H}_A)\hat{\cal H}_{\rm nano}(t)(\openone_S\otimes\hat{\tilde H}_A)$ with $\hat{\tilde H}_A=(\hat{\Sigma}_{x,A}+\hat{\Sigma}_{z,A})/\sqrt2$ the Hadamard gate for the CPB and $\hat{\Sigma}_{z,A}=\ket{a_+}\bra{a_+}-\ket{a_-}\bra{a_-}$. Assuming the same working conditions as in the main text, this changes the Hamiltonian of the system into 
\begin{equation}
\begin{aligned}
\hat{\cal H}'_{\rm nano}(t)&=\omega_S\hat b^\dag\hat b+\lambda_t(\hat b+\hat b^\dag)\otimes(\Sigma_z+\openone_A)\\
&=\omega_S\hat b^\dag\hat b+2\lambda_t(\hat b+\hat b^\dag)\otimes\ket{a_+}\bra{a_+}
\end{aligned}
\end{equation}
which is formally equivalent to $\hat{\cal H}'_{\rm micro}(t)$. We can then use the same gate decompositions discussed in details above to run the protocol for the reconstruction of the characteristic function. Needless to say, an alternative to this procedure would be to define a different gate-decomposition scheme based on the form of $\hat{\cal H}_{\rm nano}(t)$, a goal that is left for future work.   


\begin{thebibliography}{99}
\bibitem{nota} The relaxation of the assumption $Q_g(t)\sim0$ simply introduces a free term proportional to $\hat\Sigma_{x}$ that does not affect our protocol.
\end{thebibliography}
\end{document}